\documentclass[aps,prc,onecolumn,showpacs,superscriptaddress]{revtex4}
\usepackage[latin1]{inputenc}
\usepackage{bm}
\usepackage{epsfig}
\usepackage{amsthm}
\usepackage{amsfonts}
\usepackage{float}
\usepackage{amsmath,amssymb}
\usepackage{color}
\usepackage{subfig}
\usepackage{graphicx}
\usepackage{dcolumn}
\usepackage{bm}
\usepackage{hyperref}

\newcommand{\be}{\begin{equation}}
\newcommand{\ee}{\end{equation}}
\newcommand{\bea}{\begin{eqnarray}}
\newcommand{\eea}{\end{eqnarray}}

\newcommand{\vp}{\bold p}

\begin{document}

\title{Effect of magnetic screening mass on the diffusion of heavy quarks} 
\author{Mahfuzur Rahaman}
\email{mahfuzurrahaman01@gmail.com}
\affiliation{Variable Energy Cyclotron Centre, 1/AF Bidhan Nagar, Kolkata 700 064, India and Homi Bhabha National Institute, Anushaktinagar, Mumbai 400 094}
\author{Santosh K. Das}
\email{dsantoshphy@gmail.com}
\affiliation{School of Physical Sciences, Indian Institute of Technology Goa, Ponda 403401, Goa, India}
\author{Jan-e Alam}
\email{jane@vecc.gov.in}
\affiliation{Variable Energy Cyclotron Centre, 1/AF Bidhan Nagar, Kolkata 700 064, India and Homi Bhabha National Institute, Anushaktinagar, Mumbai 400 094}
\author{Sabyasachi Ghosh}
\email{sabyaphy@gmail.com}
\affiliation{Indian Institute of Technology Bhilai, GEC Campus,Sejbahar, Raipur 492015, Chhattisgarh, India}

\begin{abstract}
The drag and diffusion coefficients of heavy quarks propagating through quark gluon plasma
(QGP) have been estimated by shielding both the electric and magnetic type infra-red divergences.
The electric type screening in perturbative quantum chromodynamics (pQCD) has been
widely studied and used in evaluating the diffusion coefficient of heavy quarks (HQs).
To our knowledge the impact of magnetic screening in diffusion coefficients of HQs is not
studied before.
It is found that the effect of magnetic screening mass on the drag and diffusion coefficients
of HQs is quite significant and its contribution should not be ignored for explaining the experimental data
of heavy quark observables.
\end{abstract}
%
%
\maketitle
%
%
%
\section{Introduction}
\label{sec:intro} 
The main aim of the experimental heavy ion collision program at Relativistic Heavy Ion Collider (RHIC)
and Large Hadron Collider (LHC) is to create and characterize quark gluon plasma (QGP), a state
of matter where the properties of the system is governed by the deconfined quarks and gluons. 
Several probes have been proposed  to characterize QGP~\cite{cywong,hatsuda}. Among those signals, heavy
quarks (HQ)~\cite{Prino:2016cni, Andronic:2015wma, Rapp:2018qla, Aarts:2016hap, Dong:2019unq, Cao:2018ews} are 
considered as promising probe of QGP because:(i) HQs $(Q\bar{Q})$
are mainly created in the primary collisions of partons  as production of heavy objects in 
secondary  collisions at the late stage of the evolution becomes improbable due to lack of threshold energy
and (ii) the survivability of a $Q$ (or $\bar{Q}$)
during the entire evolution of the QGP is high compared to
the light partons because it is difficult for the 
$Q$ ($\bar{Q}$) to find a $\bar{Q}$ (or $Q$) in the bath to annihilate with, 
as they are thermally suppressed by a factor 
$\sim exp(-m/T)$  (since $T/m<<1$, where $T$ is the temperature
of the QGP and $m$ is the mass of the HQs). Therefore, the HQs witness the entire evolution
of QGP - from its creation  to hadronization and hence it is a useful tool
to probe the properties of QGP. While propagating through QGP the HQs created with high momentum 
interact with its constituents i.e. light quarks and gluons and gets slowed down as
its momentum gets diffused in the medium. The momentum diffusion of the HQs gets reflected 
through the suppression of its momentum distribution.
Two quantities: the nuclear suppression factors and the
elliptic flow ~\cite{moore,rappv2,rappprl,gossiauxv2,Das:2009vy,alberico,Uphoff:2012gb, 
Cao:2013ita,Das:2015ana,Song:2015sfa,Cao:2016gvr,Scardina:2017ipo} are capable of 
estimating the momentum diffusion coefficients of 
of HQs. Therefore, it is crucial to accurately estimate the diffusion of HQs by including
as many physical processes ~\cite{Prino:2016cni, Rapp:2018qla,Das:2015aga,Das:2012ck} as 
possible which affects the HQs in the bath. 
Several advances have been made, over the years,  to evaluate diffusion coefficient with
realistic Debye 
mass and running coupling~\cite{Rapp:2018qla,gossiauxv2,alberico,Singh:2018wps} to improve the description 
of the data.  In this
context we attempt to include the  screening of color magneto-static interaction
in the evaluation of drag and diffusion coefficients. 
 
The paper is organized as follows. In the next section we define the drag and momentum
diffusion coefficients. In section III the effects of magnetic screening on the shift in
pole of the gluon propagator is outlined. Section IV is devoted to present the results 
on drag and diffusion coefficients. The summary and conclusions is presented in section V.  

\section{Drag and momentum diffusion coefficients}
\label{sec:form}
The dynamics of heavy quarks in quark-gluon plasma (QGP) composed of
thermal light quarks and gluons at temperature $T$ can be modeled by 
Fokker-Plank (FP) equation \cite{van Kampen,Svetitsky}
\begin{equation}
\frac{\partial f}{\partial t}=\frac{\partial}{\partial p_i} \Big[A_i(p)f+\frac{\partial}{\partial p_j}[B_{ij}(p)f] \Big]
\end{equation}
where $f$ is the heavy quark distribution function and $p$ is the momentum.  The kernels,
$A_i$ and $B_{ij}$ are related to the drag and diffusion coefficients which are required 
as inputs along with the initial
and boundary conditions to solve the FP equation. The $A_i$ is defined as:
 
\begin{eqnarray}
A_i\hspace{-.1cm }& = & \hspace{-.1cm }\frac{1}{2E_p}\int\frac{d^3q}{(2\pi)^3 2E_q}\int\frac{d^3q'}{(2\pi)^3 2E_q'}\int\frac{d^3p'}{(2\pi)^3 2E_p'}\frac{1}{g_Q}\sum|M|^2 (2\pi)^4 \delta^4(p+q-p'-q')\hat{f}(q)[(p-p')_i]
\\\hspace{-.1cm }& = & \hspace{-.1cm }\langle\langle(p-p')_i \rangle\rangle \sim\,\int\, Phase space (p-p')_i|M|^2 
\end{eqnarray}
where $g_Q$ is the degeneracy factor, $p$ is the initial momentum and 
$p^\prime$ is the momentum after the scattering of the HQs with the bath particles,
$\hat{f}(q)$ is the thermal distributions of light quark or gluon. 
The quantity, $A_i$ is basically the change in momentum 
of the HQ weighted by the interaction (through $|M|^2$) {\it i.e.} it is the change in momentum due to its interaction 
with the particles in the bath, known commonly as drag.
The other kernel, $B_{ij}$ is given by
  
\begin{equation}
B_{ij}=\frac{1}{2}\langle\langle(p'-p)_i(p'-p)_j \rangle\rangle 
\end{equation}
Similar to $A_i$, the quantity $B_{ij}$ is the average change in the quadratic power of the
exchanged momentum due to its interaction with the particles in the medium, known as 
diffusion coefficient. 

The above two kernels can be decomposed as
\begin{eqnarray}
  A_i \hspace{-.1cm }& = & \hspace{-.1cm }  p_i A(p^2)\\
  B_{ij} \hspace{-.1cm }& = & \hspace{-.1cm }   \Big \{  \delta_{ij}-\frac{p_ip_j}{p^2}\Big\}  B_0(p^2)+\frac{p_ip_j}{p^2}   B_1(p^2)
\end{eqnarray}
where the coefficients, $A$, $B_0$ and $B_1$  are given by
\begin{eqnarray}
A  \hspace{-.1cm }& = & \hspace{-.1cm } p_i A_i/p^2
\\ \hspace{-.1cm }& = & \hspace{-.1cm }\langle\langle 1 \rangle\rangle -\frac{\langle\langle (\vec{p'}\cdot\vec{p} )^2 \rangle\rangle }{p^2}
\\B_{0}\hspace{-.1cm }& = & \hspace{-.1cm }  \frac{1}{2}\Big \{  \delta_{ij}-\frac{p_ip_j}{p^2}\Big\}  B_{ij}
\\ \hspace{-.1cm }& = & \hspace{-.1cm } \frac{1}{4}\Big \{  \langle\langle p'^2 \rangle\rangle -\frac{\langle\langle (\vec{p'}\cdot\vec{p} )^2 \rangle\rangle }{p^2}\Big\}
\\B_{1} \hspace{-.1cm }& = & \hspace{-.1cm }  \frac{p_ip_j}{p^2}B_{ij}
\\ \hspace{-.1cm }& = & \hspace{-.1cm } \frac{1}{2}\Big \{  \frac{\langle\langle (\vec{p'}\cdot\vec{p} )^2 \rangle\rangle }{p^2}-2  \langle\langle (\vec{p'}\cdot\vec{p} )^2 \rangle\rangle  + p^2 \langle\langle 1 \rangle\rangle \Big\}  
\label{dragdiff}
\end{eqnarray}
The drag($\gamma$) and diffusion ($D$) coefficients are defined as $A_i=\gamma p_i$ and 
$B_{ij}=D\delta_{ij}$ in the limit $p\rightarrow 0$.  
It is clear from the above expressions that we need to evaluate the 
thermal average, denoted by $<<.>>$ (weighted by the interactions) of the some
combinations of function of $\vec{p}$ and $\vec{p^\prime}$. 
In the center of momentum frame the quantity, $<<F(p^\prime)>>$ can be written as:
\begin{equation}
\langle\langle F(\vec{p'}) \rangle\rangle =\frac{1}{(2\pi)^5}\frac{1}{2 E_p}\int\frac{d^3q}{2E_p}\int d \Omega_{c.m.}\frac{s-m^2}{8s}\frac{1}{g_Q}\sum|M|^2\hat{f}(\vec{q})F(\vec{p'})
\label{eq13}
\end{equation} 
In the above equation, $\vec{p^\prime}$ can be expressed in terms of $p$ and $q$,
$s$ is the Mandelstam variable defined as $s=(E_p+E_q)^2-(\vec{p}-\vec{q})^2$. 
In Eq.~\ref{eq13} $m$ is the mass of heavy flavor. We 
have taken masses of the light flavours as zero.
$|M|^2$ is the scattering matrix elements for the possible pQCD
process as shown in Fig.~\ref{Fig1_1}. The amplitude for the processes displayed in 
Fig.~\ref{Fig1_1}a and Fig.~\ref{Fig1_1}d is infrared 
divergent. To regulate it, a mass (usually Debye screening mass\cite{LeBellac}) 
is introduced\cite{Svetitsky}, which basically rejects amplitudes at small angle scattering.

The integral in Eq.~\ref{eq13} can be written as~\cite{Svetitsky}:
\begin{equation}
\langle\langle F(\vec{p'}) \rangle\rangle =\frac{1}{1024 \pi^4}\frac{1}{E_p}\int_{0}^{\infty}qdq d(cos\chi)\frac{s-m^2}{s}\hat{f}(\vec{q})\int_{-1}^{1}d(cos\theta_{c.m.}\frac{1}{g_Q})\sum|M|^2\int_{0}^{2\pi}d \phi_{c.m.}F(\vec{p'})
\label{eq14}
\end{equation}
The integral over  $ \phi _{cm}$ is trivial and the remaining three dimensional integral can be evaluated numerically.

It is well known that the non-abelian gauge theories like QCD admits monopole like field configuration,
therefore, in a thermal system the monopole-antimonopole excitations can screen the color magnetic
interactions. In the present work we include such screening effects to estimate the drag and diffusion
coefficients.   This will affect the matrix elements of the processes shown in
Fig.~\ref{Fig1_1}a and Fig.~\ref{Fig1_1}d. 

\section{Color magneto-static screening effects of diffusion of HQs}

The inclusion of magnetic screening in evaluating the drag and diffusion coefficients perturbatively
can be achieved by shielding the infra-read divergences appears through the exchange of gluons 
in the $t$-channel process.  
It is well known in quantum field theory that the infrared divergences occur when a
massless particle is exchanged in the $t$ (or $u$) diagrams. 
These divergences can be cured by taking the  hard thermal loops 
(HTL)\cite{htl} approximations, where divergences can be cured by taking the electric screening mass,
in this sense, HTLs are 'electric type'.The HTL effective action gives a gauge-invariant definition of 
electric screening mass and  incorporate the electric screening in a systematic way in perturbative expansion. 
The perturbation also contains "magnetic type" divergences, which can be cured by taking the magnetic 
screening mass to screen the magneto-static interactions \cite{linde}. Non-Abelian gauge theories has 
thermally excited magnetic monopole like field configuration. The presence of monopole and 
antimonopole in the plasma screens the magnetic interactions similar to the
screening of  electrostatic interactions\cite{nair}.
 
 The qualitative origin of magnetic mass in high temperature gauge theory can be explained (see
\cite{nair} for example) as, in Eucliden field theory, the Matsubara frequency $\omega_n=2\pi n T$, 
the temporal component of the gauge field can be expanded as 
$A_1(x_0, \vec{x})=\sum_{n=0}^{\infty}A_{i,n}(\vec{x})e^{2\pi i T x_0}$. The modes having non-zero Matsubara frequencies will be insignificant at high temperature and  also the modes having wavelength long compared to $1/T$ do not contribute substantially. Hence, the theory comprises only the Matsubara 'zero mode' and effectively the theory  becomes a 3D  Yang-Mills theory. Yang-Mills theories in 3D (or (2+1)D in Wick rotated version)   are expected to have a mass gap, at high temperature, this mass gap represents the magnetic mass   of the (3+1)-dimensional theory \cite{Linde} whereas the non zero Matsubara modes $(\omega _n\neq0)$ will give some small correction to the magnetic mass. 
 
Self-consistent solution of  gap equation
upto one loop order \cite{mass_nair,Alexanian} leads to the following
expression for the  magnetic mass ($m_{mag}$): 
\begin{equation}
m_{mag}\approx 1.22\frac{g^2T N}{2\pi}
\end{equation}
where  $g$ is the (3+1) dimensional coupling. 
There are some ambiguities about the value of magnetic mass of gluon depending on the different methods used in the  perturbative and non-perturbative  calculations. Several perturbative calculations estimates zero magnetic mass of gluon  to an accuracy of $g^{3/2}T$\cite{Toimela}. 

Several authors obtained different value $m_{mag}$.
For example while $m_{mag}\sim g^2 T$ is obtained in~\cite{Linde,Kalashnikov1,Kalashnikov2,Jackiw,Cornwall},
$m_{mag}\sim g^{2/3}T$ and $m_{mag}=0$ has been quoted in \cite{Kapusta} and \cite{Kalashnikov} respectively
depending on different non-perturbative methods used in evaluations.
There are some non-perturbative estimations from lattice  QCD
based calculations\cite{Billoire} which gives  $m_{mag} =(0.24-0.27)g^2T$ and  $ 0.255 g^2T$\cite{Biro}.
Some recent lattice QCD based calculations of the magnetic mass of gluon gives $m_{mag}\sim (0.4-0.6)g^2T$
~\cite{Maezawa,Hart,Nakamura,Bak}. Earlier, the magnetic contribution on heavy-flavor jet energy 
loss was investigated by incorporating magnetic mass of gluons in Refs.\cite{Djordjevic1,Djordjevic2,
Djordjevic3,Djordjevic4}, where it has been shown that the incorporation of magnetic mass enhances the 
energy loss of jets propagating through dynamical QCD medium and the magnetic contribution is smaller 
than the electric contribution(i.e. only Debye mass taking into account to screen the infrared divergences) but the two types of masses contribute at $O(g^2)$.

\subsection{Debye mass upto next-to leading order(NLO)}

The value of magnetic mass of gluon vanishes at  the one-loop level  ($\mathcal{O}(g)$ )
\cite{Kapusta} and it is non-zero in NLO  
  ($\mathcal{O}(g^2)$),
as discussed above. Therefore, in order to compare the contribution of magnetic mass, we should also consider 
electric screening in NLO. 
Several  authors estimated $m_D$ beyond LO. Amongst them, Rebhan et.al. found Debye mass in $\mathcal{O}(g^2)$  
as \cite{Rebhan}
  \begin{equation}
   m_D=m_{0D}+\frac{Ng^2T}{4\pi }\ln\frac{m_{0D}}{g^2T} + c_N g^2T +\mathcal{O}(g^3)
   \end{equation}
   where $c_N$ is purely non-perturbative correction and is given by  $c_N=1.58 \pm 0.20$ for    SU(2) and
   $c_N=2.46 \pm 0.15$ for  massless quarks \cite{Kajantie_debye}. 
Defu et al~\cite{Defu} estimated the NLO contribution to $m_D$ as: 
\begin{equation}
 \delta m_D^2=m_{0D}^2 \sqrt{\frac{6}{2N+N_f}}\frac{gN}{2\pi} \Big \{  \ln\frac{ 2\sqrt{\frac{2N+N_f}{6}}+\frac{ N g}{8\pi  } (\ln\frac{1}{\alpha_s}+1)}{\frac{ N g}{8\pi  } (\ln\frac{1}{\alpha_s}+1)}-\frac{1}{2} -\frac{ \frac{ N g}{8\pi  }(\ln \frac{1}{\alpha_s}+1) }{2\sqrt{\frac{2N+N_f}{6}}}\Big \} 
\end{equation}
where $\alpha _s=g^2/4\pi$, $N_f$ is the number of flavour and $m_{0D} $ is the LO contribution of the Debye mass of SU(N) non-Abelian theory and it is given by
\begin{equation}
m_{0D}^2=\frac{g^2T^2}{3} \Big (N+\frac{N_f}{2} \Big)
\end{equation}
Therefore, the total contribution to screening mass (both electric and magnetic) upto NLO contribution is:
\begin{eqnarray}
m_D^2  \hspace{-.1cm }& = & \hspace{-.1cm }m_{0D}^2 +\delta m_D^2    \\   \hspace{-.1cm }& = & \hspace{-.1cm }\frac{g^2T^2}{3} \Big (N+\frac{N_f}{2} \Big) \Big [ 1+\sqrt{\frac{6}{2N+N_f}}\frac{gN}{2\pi} \Big \{  \ln\frac{ 2\sqrt{\frac{2N+N_f}{6}}+\frac{ N g}{8\pi  } (\ln\frac{1}{\alpha_s}+1)}{\frac{ N g}{8\pi  } (\ln\frac{1}{\alpha_s}+1)}-\frac{1}{2} -\frac{ \frac{ N g}{8\pi  }(\ln\frac{1}{\alpha_s}+1) }{2\sqrt{\frac{2N+N_f}{6}}}\Big \}\Big]
\end{eqnarray}
In this work, we will use the value of Debye mass upto NLO as given in Eq.20.
In the present work, we investigate the effect of non-zero magnetic mass of gluons on the 
drag and diffusion coefficients of HQs while it is moving through the QGP. 
We have taken the value of magnetic mass as $0.4<m_{mag}/m_D <0.6$, \cite{Maezawa,Nakamura,Hart,Bak}, 
where   $m_D$ is the Debye mass of gluons.
   
If we include the magnetic masses of the gluons, then the pole mass of gluon will be 
$\bf{m'\equiv\sqrt{m_D^2 +m_{mag}^2}}$
($m'$ is now used to regulate or screen the infrared divergence). 
This will modify the scattering amplitudes ~\cite{Combridge}
and  hence the cross-sections of the different scattering processes. 

\subsection {Scattering matrix elements}
\begin{figure}[t]
 	\centering	  
 	\includegraphics[width=17cm, height=3.2 cm]{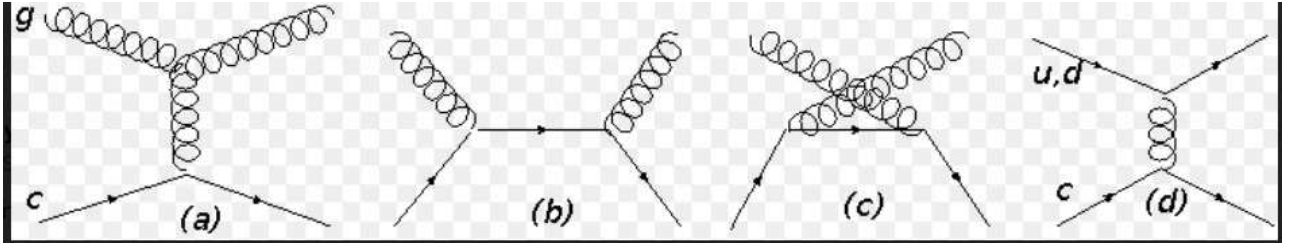} 
 	\caption{Feynman graphs for the scattering of gluon with charm quarks (a)-(c) and light quarks by charm quark} \hspace*{.2cm}  
 	\label{Fig1_1}
        \end{figure}

The inclusion of magnetic mass of gluons will change the matrix elements of diagrams 
Fig.~\ref{Fig1_1}a and Fig.~\ref{Fig1_1}d 
as these diagrams contain gluon propagator. Other diagrams will be unaffected.   The relevant
amplitudes of the processes displayed in Fig.1 are quoted below. 

\begin{eqnarray}
 \sum|M_a|^2\hspace{-.1cm }& = & \hspace{-.1cm }3072\pi^2\alpha^2_s\frac{(m^2-s)(m^2-u)}{\{t-m_D ^2(1+x^2) \}^2}
\\\sum|M_b|^2\hspace{-.1cm }& = & \hspace{-.1cm } \frac{2048}{3}\pi^2\alpha^2_s\frac{(m^2-s)(m^2-u)-2m^2(m^2+s)}{(t-s^2)^2}
\\\sum|M_c|^2\hspace{-.1cm }& = & \hspace{-.1cm } \frac{2048}{3}\pi^2\alpha^2_s\frac{(m^2-u)(m^2-s)-2m^2(m^2+u)}{(m^2-u)^2}
\\\sum M_a M_b*\hspace{-.1cm }& = & \hspace{-.1cm }\sum M_b M_a^*
\\\hspace{-.1cm }& = & \hspace{-.1cm }768 \pi^2\alpha^2_s\frac{(m^2-s)(m^2-u)+m^2(u-s)}{\{t-m_D^2(1+x^2) \}(m^2-s)}
\\\sum M_a M_c*\hspace{-.1cm }& = & \hspace{-.1cm }\sum M_c M_a^*
\\\hspace{-.1cm }& = & \hspace{-.1cm }768 \pi^2\alpha^2_s\frac{(m^2-u)(m^2-s)+m^2(s-u)}{\{t-m_D^2(1+x^2)\}(m^2-u)}
\\\sum M_b M_c*\hspace{-.1cm }& = & \hspace{-.1cm }\sum M_c M_b^*
\\\hspace{-.1cm }& = & \hspace{-.1cm }\frac{256}{3} \pi^2\alpha^2_s\frac{m^2(t-4m^2)}{(m^2-u)(m^2-s)}
\end{eqnarray}
and for light quarks ($u$ and $d$) scattering we have, 
\begin{equation}
\sum|M_d|^2=256N_f\pi^2\alpha_s^2 \frac{(m^2-s)^2+(m^2-u)^2+2m^2t}{\{t-m_D ^2(1+x^2) \}^2}
\end{equation}   
where   $ m_{mag}/m_D\equiv x $ and $0.4\,<\,x\,<\,0.6$~\cite{Djordjevic1}.

\section{Results and Discussion}
\label{sec:num}
\begin{figure}
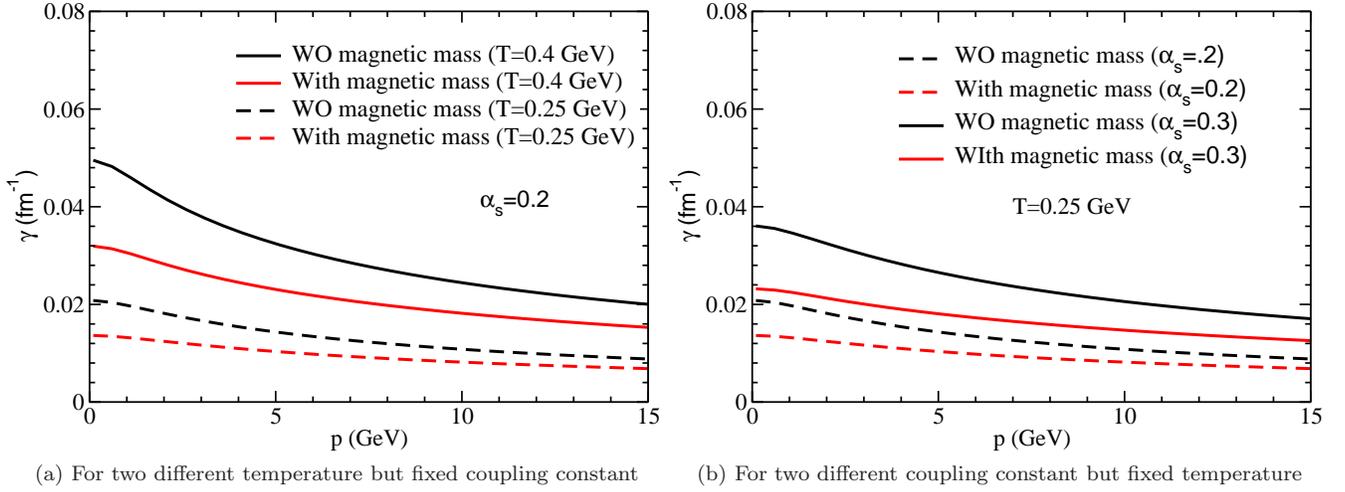
%
 	\centering
 	\subfloat[For two different temperature but fixed coupling constant]{{\includegraphics[width=8.7cm]{Fig1a.eps} }}%
 	\subfloat[For two different coupling constant but fixed temperature]{{\includegraphics[width=8.7cm]{Fig1b.eps} }}%
 	\caption{Momentum dependence of drag coefficient with and without (WO) considering magnetic mass.}%
 	\label{Fig1}%
 \end{figure}

The drag and diffusion coefficients of QGP can now be estimated by using Eqs.~\ref{dragdiff} and ~\ref{eq14}
and amplitudes taken from section III.B for various processes.  
We have taken $x=0.4$ for generating the results presented here.
In Fig.~\ref{Fig1} the variation of drag with momentum ($p$) of the heavy quark for different values
of $T$ are displayed for strong coupling, $\alpha_s=0.2$. It is clearly seen that the drag
reduces due to the inclusion of magnetic screening effects. Because of the screening of the
the forward ($\theta\rightarrow 0$) interaction cross section of the HQs with the light particles 
of the thermal bath.
The reduction in drag is more for higher $T$ (left panel of Fig.~\ref{Fig1}) and higher $\alpha_s$ 
(right panel of Fig.~\ref{Fig1}). 
Fig.~\ref{Fig2} depicts the temperature variation drag for various values of $p$ (left panel) and
$\alpha_s$ (right panel). The effects of the magnetic screening is noticeable.  The change in darg
is about $23\%$ at $T=250$ MeV around $p\sim 0$ for $\alpha_s=0.3$. 

\begin{figure}%
	\centering
	\subfloat[For two different momentum but fixed coupling constant]{{\includegraphics[width=8.7cm]{Fig2a.eps} }}%
	\subfloat[For two different coupling constant but fixed momentum]{{\includegraphics[width=8.7cm]{Fig2b.eps} }}%
	\caption{Temperature dependence of drag coefficient with and without (WO) considering magnetic mass.}%
	\label{Fig2}%
\end{figure}

 \begin{figure}
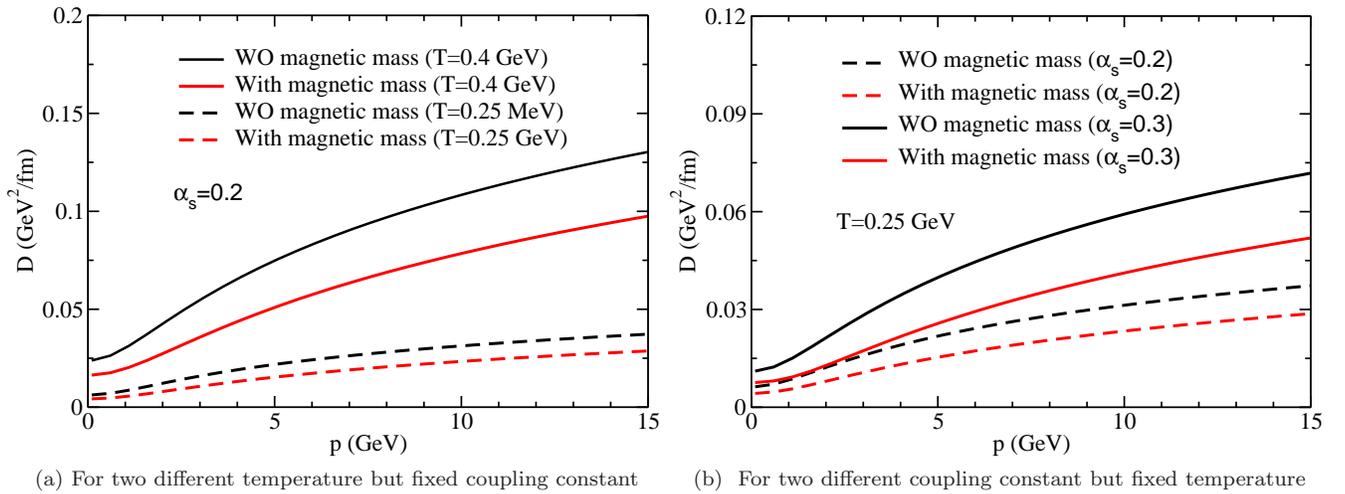
%
 	\centering
 	\subfloat[For two different temperature but fixed coupling constant ]{{\includegraphics[width=8.7cm]{Fig3a.eps} }}%
 	\subfloat[ For two different coupling constant but fixed temperature]{{\includegraphics[width=8.7cm]{Fig3b.eps} }}%
 	\caption{ Momentum dependence of diffusion coefficient with and without (WO) considering magnetic mass.}%
 	\label{Fig3}%
 \end{figure}

 	
 	

 \begin{figure}%
 	\centering
 	\subfloat[For two different momentum but fixed coupling constant ]{{\includegraphics[width=8.7cm]{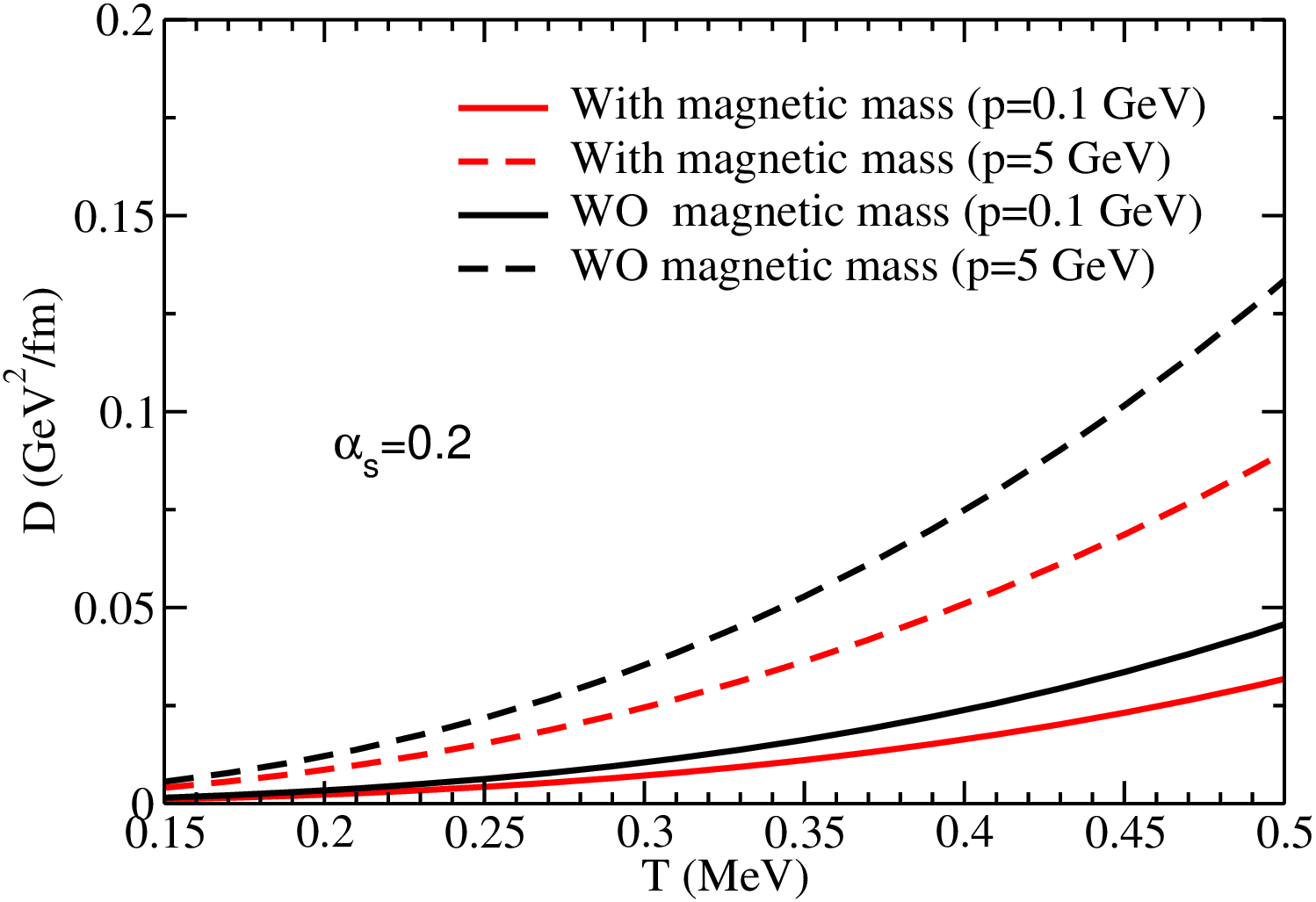} }}%
 	\subfloat[ For two different coupling constant but fixed momentum]{{\includegraphics[width=8.7cm]{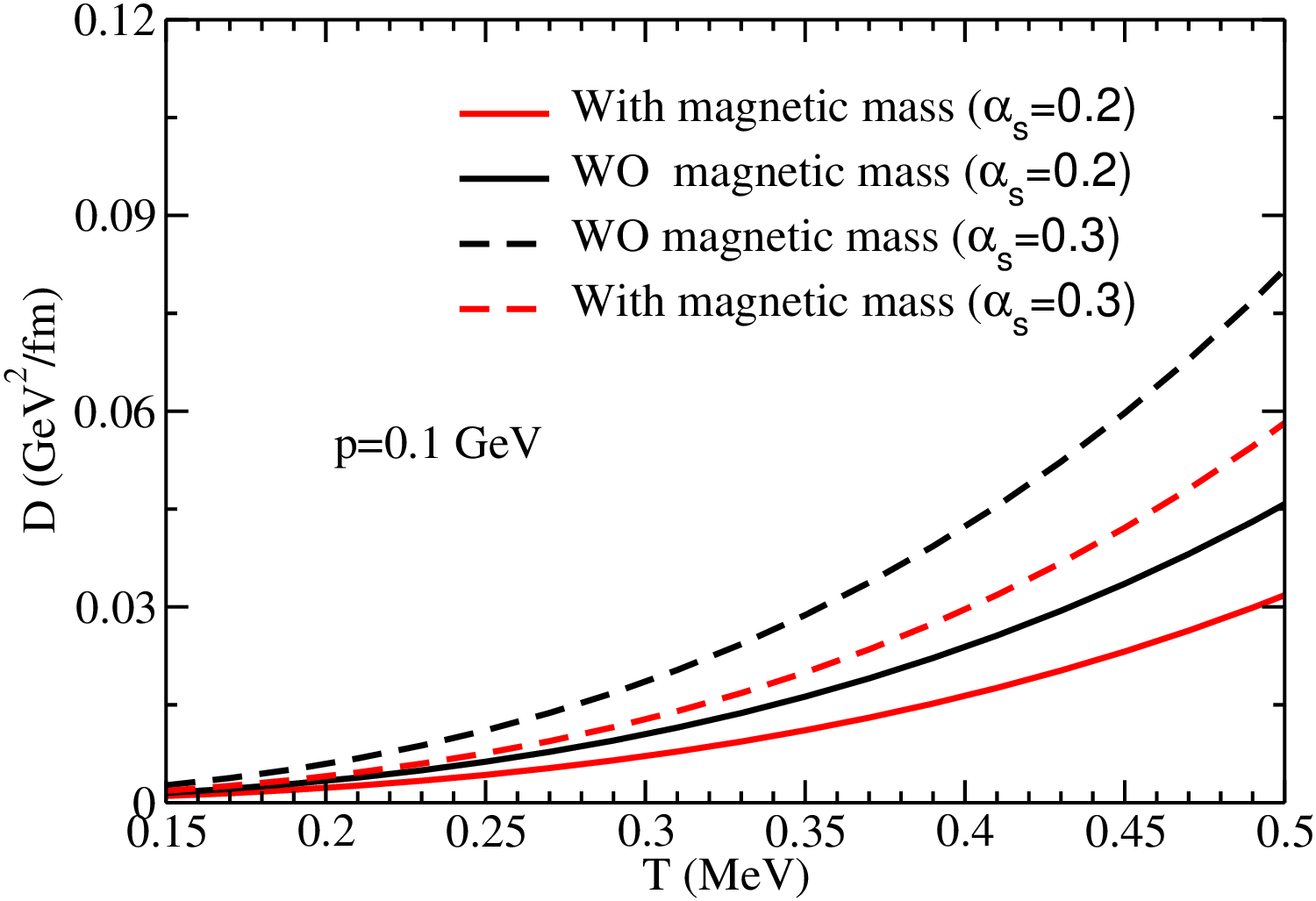} }}%
 	\caption{ Temperature dependence of drag coefficient with and without (WO) considering magnetic mass.}%
 	\label{Fig4}%
 \end{figure}

 	
 	
In Fig.~\ref{Fig3} the variation of the diffusion coefficients with $p$ is presented
for different values of $T$ (left panel) and for different values of $\alpha_s$(right panel).
Fig.~\ref{Fig4} depicts the temperature variation of diffusion coefficient for various values of $p$ (left panel) and
$\alpha_s$ (right panel). 
Similar to drag coefficient, momentum diffusion coefficient reduces due to the magnetic screening.
The reduction is more at higher $p$ and $T$ because 
at higher $T$, $\delta m_D$ is larger which suppresses the interaction cross section more.  
The momentum diffusion coefficient is a measure of the squared average momentum transfer in a collision.
In terms of spatial diffusion coefficient $D_s$ ~\cite{He:2012df,Das:2016llg}, defined as 
\be
D_s=\frac{T}{m_c\gamma(\vp=0,T)}~,
\ee
we will get enhanced $D_s$ due to magnetic screening. 
%
 %
 %

For a approximate estimation, if we express nuclear modification factor $R_{AA}\sim\,e^{-\gamma\Delta\tau}$
where $\Delta\tau$ is the duration of the QGP phase~\cite{mhe,Mitra:2014ipa}, then we can get an estimation
for percentage change in $R_{AA}$ due to magnetic screening. Considering an average drag 
$\gamma_{WO}=\frac{0.036+0.085}{2}=0.06$ fm$^{-1}$ without magnetic screening and 
$\gamma_{W}=\frac{0.023+0.053}{2}=0.038$ fm$^{-1}$ with magnetic screening for $\alpha_s=0.3$, 
${\vec p}=0$ within the temperature
range $T=0.250-0.400$ GeV, we can obtain the percentage change in $R_{AA}$ defined as 
\be
\frac{\Delta R_{AA}}{R_{AA}}=\frac{e^{-\gamma_{W}\Delta\tau}-e^{-\gamma_{WO}\Delta\tau}}{e^{-\gamma_{WO}\Delta\tau}}\times 100\%~.
\ee
Assuming $\Delta\tau=6$ fm, we get 
$\frac{\Delta R_{AA}}{R_{AA}}\approx\frac{0.79-0.69}{0.69}\times 100\%\approx 14.4\%$ enhancement
due to magnetic screening. Therefore, we conclude that magnetic screening reduces the
heavy quark suppression. The magnetic screening  
needs to be considered in theoretical calculations while analyzing experimental data 
because in dynamical QCD magnetic screening is inevitable. 
Detail calcualtion including the hydrodynamic evolution of the QGP should be done
to estimate its effect which is beyond the scope of the present work where
we have provided a back-of-the-envelop estimate.

The shear viscosity, $\eta\sim n<p>\lambda$, is interpreted as
the ability of transferring average momentum $<p>$ over a distance of magnitude
mean free path ($\lambda$) in a medium with particle density $n$. Therefore, the momentum
diffusion will be related to shear viscosity ~\cite{surasree}. Therefore, the magnetic
screening will affect the shear viscosity which can be measured through the elliptic flow
of produced particles in heavy ion collisions. 
The impact of magnetic screening on other transport coefficients like shear, bulk
viscosities, thermal and electrical conductivities of quark gluon plasma and its effects
on heavy ion phenomenology might be considered in future.

 	
 	
\section{Summary and Conclusion}
In this work we have incorporated the magnetic mass of gluons apart from the electric 
screening mass, 
which is used  to regulate the infrared divergence occurring in the processes
involving the gluon propagator in the tree level $t$ channel diagrams.  The effects of the
magnetic screening on the transport coefficients (drag and diffusion) of HQs in QGP have been 
studied. It has been observed that the magnetic screening affects the transport coefficients
significantly.   
The reduction in drag and diffusion coefficients of HQs will be reflected in the experimentally 
measured quantities like the suppression of high momentum heavy flavored hadrons
and in the elliptic flow of these hadrons. 
Within the temperature range $T=0.250-0.400$ GeV, 
our results show approximately $38-36\%$ reduction of drag coefficient of charm quark due to magnetic screening. Roughly considering the exponential suppression probability
of charm quark with an average (over the fireball life time) drag, we have found approximately $14.4\%$
enhancement of nuclear suppression factor due to magnetic screening effect. Hence, for completeness, along with
other effects, the magnetic screening should be considered while  contrasting theoretical analysis with 
experimental data of heavy quark suppression to reliably characterize QGP.
It will be interesting to see the impact of magnetic screening which is inevitable in dynamical QCD
on other transport coefficients like shear, bulk viscosities and thermal, electrical conductivities of quark gluon plasma.

{\bf Acknowledgment:} M. R. is grateful to Department of Atomic Energy, 
Government of India for financial support.
S.K.D. acknowledges the support by the National Science Foundation of China (Grants No.11805087 and No. 11875153)


\begin{thebibliography}{99}
%
\bibitem{cywong} C. Y. Wong, Introduction to high energy heavy ion collisions, World Scientific (1994).

\bibitem{hatsuda} K. Yagi, T. Hatsuda and Y. Miake, Quark Gluon Plasma,
Cambridge University Press (2005).





\bibitem{Prino:2016cni}
  F.~Prino and R.~Rapp,
  J.\ Phys.\ G {\bf 43}, no. 9, 093002 (2016)


  
  

\bibitem{Andronic:2015wma}
  A.~Andronic {\it et al.},
  Eur.\ Phys.\ J.\ C {\bf 76}, no. 3, 107 (2016)

\bibitem{Rapp:2018qla}
  R.~Rapp {\it et al.},
  Nucl.\ Phys.\ A {\bf 979}, 21 (2018)


\bibitem{Aarts:2016hap}
  G.~Aarts {\it et al.},
  Eur.\ Phys.\ J.\ A {\bf 53}, no. 5, 93 (2017)



\bibitem{Dong:2019unq}
  X.~Dong and V.~Greco,
  Prog.\ Part.\ Nucl.\ Phys.\  {\bf 104}, 97 (2019).


\bibitem{Cao:2018ews}
  S.~Cao {\it et al.},
  Phys.\ Rev.\ C {\bf 99}, no. 5, 054907 (2019)


  
\bibitem{moore} G. D. Moore, D Teaney, Phys. Rev. C {\bf 71}, 064904 (2005)

\bibitem{rappv2} H. van Hees, V. Greco and R. Rapp, Phys. Rev. C {\bf 73}, 034913 (2006)
  
\bibitem{rappprl} H. Van Hees, M. Mannarelli, V. Greco and R. Rapp, Phys. Rev. Lett. {\bf 100},192301 (2008) 
  
\bibitem{gossiauxv2} P. B. Gossiaux, J. Aichelin,  Phys. Rev. C {\bf 78} 014904 (2008); P.~B.~Gossiaux, J.~Aichelin, T.~Gousset and V.~Guiho, J.\ Phys.\ G {\bf 37}, 094019 (2010)


\bibitem{Das:2009vy} 
S.~K.~Das, J.~Alam and P.~Mohanty, Phys.\ Rev.\ C {\bf 80}, 054916 (2009);
S.~Mazumder, T.~Bhattacharyya, J.~e.~Alam and S.~K.~Das, Phys.\ Rev.\ C {\bf 84}, 044901 (2011)
  
  
\bibitem{alberico} W. M. Alberico {\it et al.}, Eur. Phys. J. C, {\bf 71} 1666 (2011)


\bibitem{Uphoff:2012gb} 
  J.~Uphoff, O.~Fochler, Z.~Xu and C.~Greiner,
  Phys.\ Lett.\ B {\bf 717}, 430 (2012)
  [arXiv:1205.4945 [hep-ph]].

\bibitem{Cao:2013ita} 
  S.~Cao, G.~Y.~Qin and S.~A.~Bass,
  Phys.\ Rev.\ C {\bf 88}, 044907 (2013)
 
\bibitem{Das:2015ana}
  S.~K.~Das, F.~Scardina, S.~Plumari and V.~Greco,
  Phys.\ Lett.\ B {\bf 747}, 260 (2015)
  
\bibitem{Song:2015sfa} 
  T.~Song, H.~Berrehrah, D.~Cabrera, J.~M.~Torres-Rincon, L.~Tolos, W.~Cassing and E.~Bratkovskaya,
  Phys.\ Rev.\ C {\bf 92}, no. 1, 014910 (2015)


\bibitem{Cao:2016gvr} 
  S.~Cao, T.~Luo, G.~Y.~Qin and X.~N.~Wang,
  Phys.\ Rev.\ C {\bf 94}, no. 1, 014909 (2016)
  
  
\bibitem{Scardina:2017ipo} 
  F.~Scardina, S.~K.~Das, V.~Minissale, S.~Plumari and V.~Greco,
  Phys.\ Rev.\ C {\bf 96}, no. 4, 044905 (2017)
  
  
  
\bibitem{Das:2012ck} 
  S.~K.~Das, V.~Chandra and J.~Alam,
  J.\ Phys.\ G {\bf 41}, 015102 (2013)
  
  
\bibitem{Das:2015aga} 
  S.~K.~Das, M.~Ruggieri, S.~Mazumder, V.~Greco and J.~e.~Alam,
  J.\ Phys.\ G {\bf 42}, no. 9, 095108 (2015)
  
  
\bibitem{Singh:2018wps} 
  B.~Singh, A.~Abhishek, S.~K.~Das and H.~Mishra,
  Phys.\ Rev.\ D {\bf 100}, no. 11, 114019 (2019)

\bibitem{van Kampen}N. G. van Kampen, Stochastic Processes in Physics and Chemistry (North-Holland,Amsterdam, 1981),
H. Risken,  The Fokker-Planck Equation: Methods of Solutions and Applications, 2nd edition, Springer Series in Synergetics, Springer.

\bibitem{LeBellac}  M. Le Bellac, Thermal Field Theory, Cambridge University Press (1996).
 
\bibitem{Svetitsky}B. Svetitsky, Phys. Rev. {\bf D 37} (1988) 2484.


\bibitem{htl} R. Pisarski, Physica A158, 246 (1989); Phys.Rev.Lett.
63, 1129 (1989); E. Braaten and R. Pisarski, Phys.Rev.D42, 2156 (1990); Nucl.Phys. B337, 569 (1990); ibid. B339, 310 (1990); Phys.Rev. D45, 1827 (1992); J.Frenkel and J.C. Taylor, Nucl.Phys. B334, 199 (1990);
J.C. Taylor and S.M.H. Wong, Nucl.Phys. B346, 115 (1990); R. Efraty and V.P. Nair, Phys.Rev.Lett. 68, 2891 (1992); Phys.Rev. D47, 5601 (1993); R. Jackiw and V.P. Nair, Phys.Rev. D48 (1993) 4991; J.P. Blaizot and
E. Iancu, Phys.Rev.Lett. 70, 3376 (1993); Nucl.Phys. B417, 608 (1994).
\bibitem{linde}A.D. Linde, Phys.Lett. B96 (1980) 289; D. Gross, R. Pisarski and L. Yaffe, Rev.Mod.Phys. 53 (1981) 43 

\bibitem{nair}V.P. Nair	arXiv:hep-th/9809086

\bibitem{Linde}A.D. Linde, Phys.Lett. B96 (1980) 289; D. Gross, R. Pisarski and L. Yaffe, Rev.Mod.Phys. 53 (1981) 43;A.D. Linde, Confinement of monopoles at high temperatures, Lebedev Phys. Inst. preprint (1980). 

 
\bibitem{mass_nair}V.P. Nair, Phys. Lett. B352, 117 (1995).
 
\bibitem{Alexanian}G. Alexanian and V.P. Nair,  Phys.Lett. B352,

\bibitem{Toimela} T. Toimela, Preprint HU-TET82-37, Helsinki, 9182; T. Furusawa and K. Kikkawa, Preprint OUHET 52, Osaka University.

\bibitem{Kalashnikov1} O. K, Kalashnikov, JETP Lett. 33, 165(1981). 
\bibitem{Kalashnikov2} O. K, Kalashnikov, Preprint HU-TFT 82-58, Helsinki University.
\bibitem{Jackiw}R. Jackiw, S. Y.  Pi, Phys. Lett. B 368 131 
\bibitem{Cornwall}J. M.  Cornwall, Phys. Rev. D 57 3694?700

\bibitem{Kapusta}K. Kajantie and J. Kapusta, Phys. Lett. B 110, 229 (1982), CERN preprint TH-2382, 1982. 

\bibitem{Kalashnikov} O. K, Kalashnikov, JETP Lett.39, 7(1984). 



\bibitem{Billoire} A. Billoire, G. Lazarides and Q. Shafi, Phys. Lett. B103 (1981) 450;
T.A. DeGrand and D. Toussaint, Phys. Rev. D25 (1982) 526;
G. Lazarides and S. Sarantakos, Phys. Rev. D31 (1985) 389 

\bibitem{Biro}T.S. Biro and B. Muller, Nucl.Phys. A 561 (1993) 477 

\bibitem{Maezawa}Yu. Maezawa, et al., WHOT-QCD Collaboration, Phys. Rev. D 81 (2010) 091501;

Yu. Maezawa, et al., WHOT-QCD Collaboration, PoS Lattice (2008) 194

\bibitem{Hart}A. Hart, M. Laine, O. Philipsen, Nucl. Phys. B 586 (2000) 443.

\bibitem{Nakamura}A. Nakamura, T. Saito, S. Sakai, Phys. Rev. D 69 (2004) 014506. 

\bibitem{Bak}  D. Bak, A. Karch, L.G. Yaffe, JHEP 0708 (2007) 049.
\bibitem{Djordjevic1}M. Djordjevic , M  Djordjevic  Phys. Lett. B 709 (2012) 229-233 
\bibitem{Djordjevic2}M. Djordjevic Nucl. Phys. A 00 (2018), arXiv:1209.0198
\bibitem{Djordjevic3} M. Djordjevic, J. Phys. G: Nucl. Part. Phys. 39 (2012)
\bibitem{Djordjevic4}M. Djordjevic Phys. Rev. C, 85, 034904 (2012)
\bibitem{Rebhan}A. Rebhan, Phys. Rev. D 48 (1993)  3967;A. K. Rebhan, Nucl. Phys. B430, 319; E. Braaten, A.Nieto Phys. Rev. Lett. 73 (1994) 2402
\bibitem{Kajantie_debye}K. Kajantie, M. Laine, J. Peisa, A. Rajantie, K. Rummukainen, and M. Shaposhnikov
Phys. Rev. Lett. 79, 3130
\bibitem{Defu}H. Defu, L. Jiarong,  Z. Phys. C - Particles and Fields (1996) 71, 503 

\bibitem{Combridge} L. Combridge, Nucl. Phys. B {\bf 151},429 (1979). 

\bibitem{He:2012df} 
  M.~He, R.~J.~Fries and R.~Rapp,
  Phys.\ Rev.\ Lett.\  {\bf 110}, no. 11, 112301 (2013)
  
  
\bibitem{Das:2016llg} 
  S.~K.~Das, J.~M.~Torres-Rincon, L.~Tolos, V.~Minissale, F.~Scardina and V.~Greco,
  Phys.\ Rev.\ D {\bf 94}, no. 11, 114039 (2016)



\bibitem{mhe} M. He, R. J. Fries and R. Rapp, Phys. Lett. B {\bf 701}, 445 (2011).

\bibitem{Mitra:2014ipa} 
  S.~Mitra, S.~Ghosh, S.~K.~Das, S.~Sarkar and J.~e.~Alam,
  Nucl.\ Phys.\ A {\bf 951}, 75 (2016)
 


\bibitem{surasree} S. Mazumder, T. Bhattacharyya and J. Alam, Phys. Rev. D {\bf 89}, 014002 (2014).




\end{thebibliography}
\end{document}